\documentclass{aa}
\usepackage{graphicx}
\usepackage{txfonts}
\usepackage{natbib}
\bibpunct{(}{)}{;}{a}{}{,}

\newcommand{\A}{$\alpha\ $}
\begin{document}
\title{The red and blue galaxy populations in the GOODS field: evidence for an excess of red dwarfs}

   \author{S. Salimbeni \inst{1}
   \and
   E. Giallongo \inst{1}
   \and
   N. Menci \inst{1}
   \and
   M. Castellano \inst{2}
   \and
   A. Fontana \inst{1}
   \and
   A. Grazian \inst{1}
   \and
   L. Pentericci \inst{1}
   \and
   D. Trevese \inst{2}
   \and
   S. Cristiani \inst{3}
   \and
   M. Nonino \inst{3}
   \and
   E. Vanzella \inst{3}
   }

   \offprints{S. Salimbeni, \email{salimbeni@mporzio.astro.it}}

\institute{INAF - Osservatorio Astronomico di Roma, Via Frascati 33,
I--00040 Monteporzio (RM), Italy \and Dipartimento di Fisica,
Universit\'{a} di Roma ``La Sapienza'', P.le A. Moro 2, 00185 Roma,
Italy \and INAF - Osservatorio Astronomico di Trieste, Via G.B.
Tiepolo 11, 34131 Trieste, Italy}

   \date{Received .... ; accepted ....}
   \titlerunning{GOODS-MUSIC: The red and blue galaxy populations}

  \abstract
{}
{  
We study the evolution of the galaxy population up to $z\sim 3$ as a
function of its colour properties. In particular, luminosity
functions and luminosity densities have been derived as a function
of redshift for the blue/late and red/early populations.}
{We use data from the GOODS-MUSIC catalogue which have typical
magnitude limits $z_{850}\leq 26$ and $Ks\leq 23.5$ for most of the
sample. About 8\% of the galaxies have spectroscopic redshifts; the
remaining have well calibrated photometric redshifts derived from
the extremely wide multi-wavelength coverage in 14 bands (from the U
band to the Spitzer $8 \mu$m band). We have derived a catalogue of
galaxies complete in rest-frame B-band, which has been divided in
two subsamples according to their rest-frame U-V colour (or derived
specific star formation rate, SSFR) properties.}
{We confirm a bimodality in the U-V colour and SSFR of the galaxy
sample up to $z\sim 3$. This bimodality is used to compute the LFs
of the blue/late and red/early subsamples. The LFs of the blue/late
and total samples are well represented by steep Schechter functions
evolving in luminosity with increasing redshifts. The volume density
of the LFs of the red/early populations decreases with increasing
redshift. The shape of the red/early LFs shows an excess of faint
red dwarfs with respect to the extrapolation of a flat Schechter
function and can be represented by the sum of two Schechter
functions. Our model for galaxy formation in the hierarchical
clustering scenario, which also includes external feedback due to a
diffuse UV background, shows a general broad agreement with the LFs
of both populations, the larger discrepancies being present at the
faint end for the red population. Hints on the nature of the red
dwarf population are given on the basis of their stellar mass and
spatial distributions.}
{}


\maketitle
%

\section{INTRODUCTION}

The evolution of galaxy Luminosity Function (LF) is one of the main
tools to study the structure evolution through the cosmic time. The
advent of large surveys has allowed the analysis of sub-samples of
galaxies selected as a function of their morphological,
spectroscopic or colour properties \citep[][elsewhere
G05]{strateva2001,norberg2002,madgwick2002,wolf2003,willmer,bell2004,baldry2004,hogg2004,weiner2005,blanton2005,ilbert2006,marchesini2006,baldry2006,driver2006,cucciati06,cirasuolo2006,arnouts2007,giallo2005}.
In fact these kind of studies allow us to probe the evolution of
galaxies having different star formation histories.

Of special interest are the studies concerning the statistical
properties of galaxies selected on the basis of their intrinsic
colour distribution. This distribution appears bimodal up to $z\sim
2-3$ \citep{baldry2004,blanton2005,giallo2005} and separates the
galaxies in two populations, red early types vs. blue late types. It
has been shown that this spectral classification is roughly
consistent with the correspondent morphological classification
(bulge vs. disc dominated) at least at low and intermediate
redshifts \citep{strateva2001,weiner2005}.

This bimodal colour distribution can find a natural explanation in
hierarchical models for galaxy formation \citep{menci2005,menci2006}
where two distinct populations arise in the colour distribution
based on two different star formation histories affected by the
feedback effects produced by the SN and AGN activities
\citep{menci2006}. However the effect of environmental density on
the paths of galaxy evolution can have a fundamental role. In this
context it is not clear whether the evolutionary history of galaxies
is originated by a {\it nurture} senario (galaxy properties are
affected by environment through physical mechanisms acting on
galaxies) or by a {\it nature} scenario  \citep[the evolution is
driven by the initial condition established during the formation
epoch of galaxies, e.g. ][]{mateus2007,cooper2007}.

Recent studies have estimated the shape and evolution of the LF of
galaxies selected according to their bimodal colour distribution
using both the large local Sloan survey
\citep{baldry2004,blanton2005}, and other surveys at intermediate
and high redshifts
\citep{bell2003,giallo2005,faber,willmer,ilbert2006}. Their results
show the red LF evolving mildly in density up to $z \sim 1$ with a
quite flat shape at the faint-end, although the evaluation of the
faint-end slope of the red LF remains an open issue especially at
intermediate and high redshifts where the present surveys do not
constrain the faint slope very well \citep{faber,bell2003}.

In G05 we studied the red and blue LFs, using the properties of
bimodality in colour and in specific star formation rate (SSFR),
with a complete but relatively small sample of galaxies selected in
the rest-frame B-band from low to high redshifts. We showed that the
bimodality extends at least up to $z\sim 2.5$. We also found that
the red/early galaxies decrease in their luminosity density by a
factor $\sim 5-6$ from $z\sim 0.5 $ to $z\sim 2.5-3$ in broad
agreement with the hierarchical cold dark matter model. These
results provided a first picture of the evolution of the red and
blue LFs up to high redshifts relaying on a relatively deep but
small sample. For a more reliable picture a wider sample at high
redshift is clearly needed. For this purpose larger areas with deep
near-IR imaging are required.

Thanks to the wide area ($\sim 140\ arcmin^2$) and to the deep
near-IR observations, the GOODS-South survey provides a good
starting point for the study of the galaxy properties at high
redshift. In particular, the inclusion of the deep IR observations
obtained with the Spitzer telescope represent a useful constraint
for the estimate of the physical properties of galaxies at high
redshift. Last but not least the extensive spectroscopic follow up
obtained in this field provides a wide set of spectroscopic
redshifts. From this public data set we have obtained a multi-colour
catalogue of galaxies we named GOODS-MUSIC \citep[GOODS MUlticolour
South Infrared Catalog, ][]{grazian}. This catalog, where galaxies
are selected both in the $z_{850}$ and $Ks$ bands, contains
information in 14 bands from the U to the Spitzer 8 $\mu m$ band,
and all the available spectroscopic information. For all the objects
without spectroscopic information we have obtained well calibrated
photometric redshifts by means of our photometric redshift code
\citep{fontana00}.

The GOODS-MUSIC catalogue has been already used to investigate the
physical and clustering properties of high redshift galaxies
\citep{fontana06,grazian2006_eros,grazianc,pentericci2007,castellano2007}.
With the present paper we study the galaxy LFs of the red and blue
populations, enlightening evolutionary features which are
characteristic of the two populations.

The paper is organised as follows: in section 2, we describe the
basic features of our dataset. In section 3, we describe the
bimodality properties of the sample and we define the loci of
minimum for the selection of the red/early and the blue/late
sub-samples as a function of $z$. In section 4, we compute the shape
and evolutionary properties of the LFs and the luminosity density of
both populations. Section 5 is devoted to the analysis of the
physical properties of the faint early population.

All the magnitudes used in the present paper are in the AB system.
An $\Omega_\Lambda=0.7$, $\Omega_M=0.3$, and $H_0=70$ km s$^{-1}$
Mpc$^{-1}$ cosmology is adopted.

\section{THE GOODS-MUSIC SAMPLE}

We use the multicolour catalogue extracted from the southern field
of the GOODS survey, located in the Chandra Deep Field South. The
procedure adopted to extract the catalogue is described in detail in
\cite{grazian}. Here we summarise the general features.

The photometric catalogue was obtained combining 14 images from the
U-band up to 8 $\mu m$. More specifically it includes two $U$ band
images from the ESO 2.2 m telescope, an $U$ image from VLT-VIMOS,
the ACS-HST images in four bands $B$, $V$, $I$ and $z_{850}$, the
VLT-ISAAC $J$,$H$ and $Ks$ bands and the Spitzer bands at 3.6, 4.5,
5.8 and 8 $\mu m$. All the images analysed have an area of 143.2
$arcmin^2$, except for the U-VIMOS image (90.2 $arcmin^2$) and the H
image (78.0 $arcmin^2$). The multicolour catalogue contains 14847
objects, selected either in the $z$ and/or in the $Ks$ band (version
(1.0)). As in previous papers \citep[][ G05]{poli2003} we select
galaxies in different bands depending on the redshift interval; more
specifically we select galaxies in the $z$ band at low redshifts
(0.2-1.1) and in the $Ks$ band at higher redshifts (1.1-3.5).  This
allows us, as explained below, to estimate the galaxy luminosity
function in the rest frame 4400 \AA \ in the overall redshift
interval (0.2-3.5). As stated by \cite{cameron2007} \citep[see
also][]{trujillo2006} a careful analysis of the selection effects
due to the detection completeness is needed. This issue is discussed
in the paper by \cite{grazian}. In that paper we have evaluated,
using simulations in the $z_{850}$ and $Ks$ bands, a 90\%
completeness level for elliptical and spiral galaxies of different
half-light radii and bulge/disk ratios. Since the depth of the image
used for the galaxy selection varies across the area, a single
magnitude limit cannot be defined in each band. As a consequence we
have divided the $z$-selected sample and the $Ks$-selected sample in
six independent catalogs, each with a well defined area and
magnitude limit, relative to a 90\% completeness level. The
$z$-selected catalogs have magnitude limits in the range
24.65-26.18, while the magnitude limits in $Ks$-selected sample
range from 21.60 to 23.80, but for most of the sample the typical
magnitudes limits are $z_{850} \sim 26.18$ and $Ks\sim 23.5$.

In summary, the $z$-selected sample has 9862 (after removing AGNs
and galactic stars) galaxies with about 10\% having spectroscopic
redshift, while the $Ks$-selected sample has 2931 galaxies with
about 27\% having spectroscopic redshifts. For the galaxies without
spectroscopic redshift we use the photometric one. Our photometric
redshift technique has been described in \cite{giallongo98} and
\cite{fontana00}. We adopt a standard $\chi^2$ minimisation over a
large set of templates obtained from synthetic spectral models; in
particular we use those obtained with P\'{E}GASE 2.0
\citep{fioc1997} as described in \cite{grazian}. The comparison with
the spectroscopic subsample shows that the accuracy of the
photometric estimation is very good, with $<|\Delta z/(1+z)|>=0.045$
in the redshift interval $0<z<6$.

As in \cite{poli2003} and \cite{giallo2005} great care was given to
the selection of the sample suited for the estimate of the
Luminosity Function. Indeed we used the $z$-selected sample for
galaxies with $z<1$ where the 4400\ {\AA}\ rest-frame wavelength is
within or shorter than the $z_{850}$-band. For this reason we
included in our LF only galaxies with $m[4400(1+z)]\le z_{AB}(lim)$.
This selection guarantees a completeness of the LF sample at $z<1$
independently of the galaxy colour although some galaxies from the
original $z$-limited sample are excluded since they have a red
spectrum, and consequently a magnitude $m[4400(1+z)]$ fainter than
our adopted threshold. The same procedure was adopted at higher
(z=1.0-3.5) redshifts using the $Ks$-selected sample. The sample
selected as described above was adopted for all the analysis
presented in this paper.

The method adopted to estimate the rest-frame magnitude and the
other physical parameters is described in previous papers
\citep{poli2003,giallo2005,fontana06}. Briefly, it is based on a set
of templates, computed with standard spectral synthesis models
\citep{bruzual2003}, chosen to broadly encompass the variety of
star--formation histories, metallicities and extinctions of real
galaxies.  To provide a comparison with previous works, we have used
the Salpeter IMF, ranging over a set of metallicities (from $Z=0.02
Z_\odot$ to $Z=2.5 Z_\odot$) and dust extinctions ($0<E(B-V)<1.1$,
with a Calzetti extinction curve). Details are given in Table 1 of
\cite{fontana2004}. For each model of this grid, we have computed
the expected magnitudes in our filter set, and found the
best--fitting template with a standard $\chi^2$ minimisation, fixing
the redshift to the spectroscopic or to the photometric one. The
best--fit parameters of the galaxy were found after scaling to the
actual luminosity of each observed galaxy. Uncertainties in this
procedure produced, on average, small errors ($\le 10 \%$) in the
rest-frame luminosity \citep{ellis1997,pozzetti2003}. Moreover, the
inclusion of the 4 Spitzer bands, longward of 2.2 $\mu$m,for
galaxies at $z>2$,  was essential to sample the spectral
distribution in the rest-frame optical and near-IR bands, and to
provide reliable constraints on the stellar mass and dust estimation
\citep[for a detailed analysis see][]{fontana06}.

\section{The semi-analytical model}\label{sec:model}

In order to make a comparison with current  hierarchical models of
galaxy formation we used our semi-analytical  model (SAM), described
here briefly \citep[for a detailed description
see][]{menci2005,menci2006}.

The model connects i) the processes related to the gas physics
(emission, radiative cooling, heating), ii) the star formation
activity (whose rate is assumed to proceed from the conversion of
the cold gas mass on a timescale proportional to the disk dynamical
timescale) and iii) the consequent Supernovae heating of the gas to
the merging histories of dark matter haloes. The model also includes
the effect of starbursts triggered by galaxy interactions and the
accretion onto supermassive black holes at the centre of galaxies
powering the AGN activity, with the corresponding energy feedback
onto the interstellar medium.

We adopt the same choice for the model free parameters (the
normalisation of the star formation timescale and of the Supernovae
energy feedback) as in the above papers; the only changes concern
the use of merger trees with a larger mass resolution ($M=10^{9}
\,M_{\odot}$) for progenitors of large mass ($M>10^{14}\,M_{\odot}$)
haloes, and the complete depletion of gas in haloes with a virial
temperature lower than $4 \cdot 10^4$ K, due to the effect of the UV
background  \citep[see also][]{somerville1999}.

\section{BIMODALITY}\label{sec:bimodality}

\subsection{Colour and SSFR properties of the GOODS-MUSIC sample}

\begin{figure}
\resizebox{\hsize}{!}{\includegraphics[width=9cm]{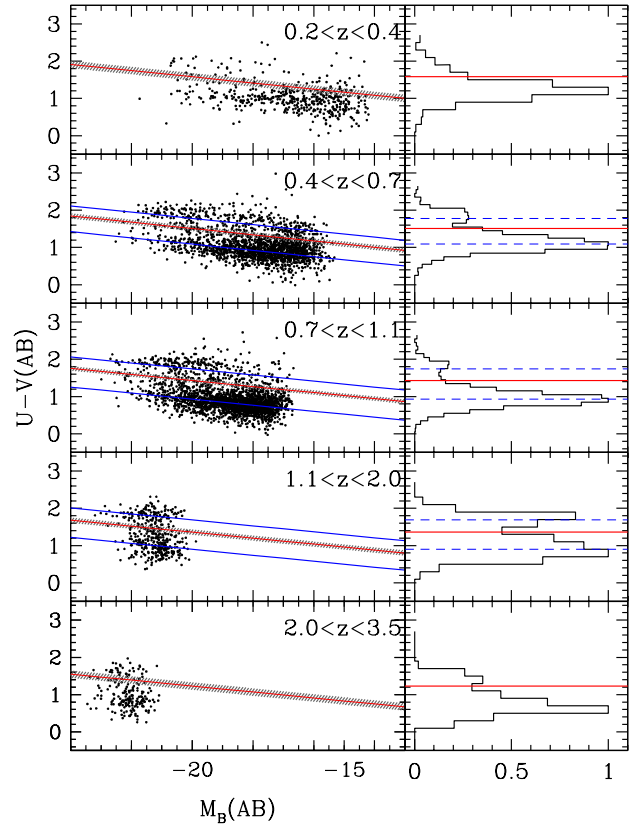}}
\caption{left panel: colour-magnitude diagrams in various redshift
intervals; the lines represent the best-fit relations for the blue
and red populations and the locus of the minimum, the shaded area is
the uncertainty on the minimum. Right panel: the histograms of
colour distribution projected at $M_B=-20$ along the best-fit
correlations, the continuous horizontal lines are the colour
separation at this magnitude, and the dash horizontal lines are the
loci of the maxima.} \label{fig:istocol}
\end{figure}

\begin{figure}[htb]
\resizebox{\hsize}{!}{\includegraphics{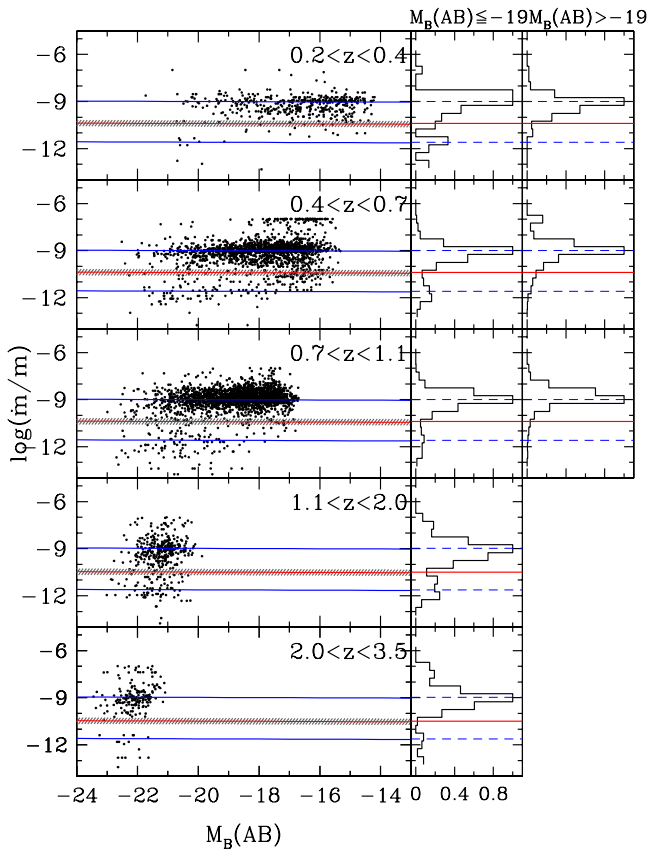}} \caption{as in Fig.
\ref{fig:istocol} but for the SSFR-magnitude distribution.}
\label{fig:istomp}
\end{figure}

The recent analysis of the spectral properties of galaxies selected
in large or deep surveys has shown the presence of a strong
bimodality in their colour distribution
\citep[e.g.][]{strateva2001,baldry2004,willmer}, allowing the
identification of two main populations, red/early and blue/late
galaxies mainly on the basis of a single colour, e.g. the rest-frame
U-V. The local distribution has been studied by \cite{strateva2001}
and \cite{baldry2004} in the framework of the Sloan survey and, at
intermediate and high redshifts, by several authors
\citep{bell2004,giallo2005,weiner2005,cirasuolo2006,cucciati06,Franceschini}.
Some effort has been devoted in explaining the observed bimodality
in the framework of the hierarchical clustering picture
\citep{menci2005,menci2006,dekel}. In particular, \cite{menci2005}
proposed  that the colour bimodality arises because of two natural
features: the star formation histories of the massive red galaxies,
which are formed in biased high-density regions, are peaked at
higher $z$ as compared to lower mass galaxies; and the existence of
a non-gravitational mass scale ($m_0$). For $m<m_0$ the star
formation is self regulated and the cold gas content is continuously
depleted by SN feedback, for $m>m_0$ the cold gas is not effectively
reheated and so the rapid cooling takes place at high-$z$. These
different evolutionary paths led to the present day red and blue
populations \citep{menci2005}. When the energy injection from AGN
feedback is included \citep{menci2006}, the bimodal distribution
appears at even higher redshifts ($z>2$).

Using the rest-frame colour we can separate the red population from
the blue to analyse the evolution of the LFs. A recent analysis of
the morphological structure of a fraction of the GOODS sample has
shown a good correlation between the red colour and the spheroidal
morphology of galaxies up to $z\sim 1.5$
\citep[see][]{Franceschini}. Moreover, as in G05 we are interested
in the galaxy evolution as a function of the star-formation
activity. In this respect, the use of the colour criterion
introduces some population mixing for the red galaxies since it is
not possible to distinguish an early-type galaxy from a dusty
star-burst using only the rest-frame $U-V$ colour. Therefore, we
have used the \cite{bruzual2003} spectral best fit of the individual
galaxy SEDs to derive the specific star formation rate (SSFR) $\dot
m_* /m_*$ (as described in the previous section). We are aware that
the absolute values in the $\dot m_*/m_*$ distribution are subject
to uncertainties due, for example, to the estimate of the dust
attenuation which depends on the extinction curve adopted and to the
methods adopted for the mass estimate. We refer to our previous
paper G05 and references therein for a description of the method
used and its reliability. However, although some degeneracy still
remains, we can use this property to separate our sample, removing
the obvious star-burst galaxies from the locus of early-type.

The results about the colour bimodality at high redshift from G05
are here confirmed at a higher statistical level. The
colour-luminosity relation is shown in Fig. \ref{fig:istocol}, while
the analogous distribution in SSFR is shown in Fig.
\ref{fig:istomp}.

The minima in the colour-magnitude distribution and in the SSFR
-magnitude distribution are used to divide the sample in red/early
and blue/late populations. For an evaluation of this relation we
have adopted the same procedure as in G05. We have fit the
distribution shown in Fig. \ref{fig:istocol} and \ref{fig:istomp}
with the sum of two gaussians whose mean is a linear function of the
absolute magnitude $M_B$. Each gaussian has a constant dispersion
and each sub-sample of galaxies, with a different magnitude limit,
has been weighted with its covering sky area. Since the statistics
of the red population is still poor, we have adopted, as in G05, the
same dependence on the absolute magnitude for both the blue and the
red populations. Taking into account the different normalisations of
the two gaussian distributions we have then derived the locus of the
formal minimum in the sum of the two gaussians, separating in this
way the two populations. The loci of the red/early and blue/late
populations are shown in Tab. \ref{tab:max}, and the minima in Tab.
\ref{tab:min}.

The resulting colour distribution projected at $M_B=-20$ along the
best-fit correlation is also displayed in Fig. \ref{fig:istocol},
with the vertical line indicating the colour separation at that
magnitude.

The same is shown for the SSFR distribution in Fig. \ref{fig:istomp}
where the relative numbers of early and late type galaxies can also
be derived in two different ranges of absolute magnitudes.

The uncertainty associated with the selection of the minima has been
derived reproducing 100 colour-magnitude plots with a MonteCarlo
analysis using 100 galaxy catalogs. In each catalogue we assigned to
each object a different redshift drawn from its probability
distribution and we associated their rest-frame absolute magnitudes
and SSFRs. The $z$ probability distribution naturally takes into
account the photometric errors and the model degeneracy in the
spectral libraries. The uncertainty region is shown in figs.
\ref{fig:istocol}, \ref{fig:istomp} as a shaded area, its value is
$\leq 0.1$ for the minima in colour, and $\sim 0.2$ for those in
SSFR.

The colour/SSFR-magnitude relations for the loci of the maxima and
minima follow the linear relations $<U-V>=\alpha \cdot (M_B+20)+
<U-V>_{20}$ and $<SSFR>=\alpha \cdot (M_B+20)+ <SSFR>_{20}$, whose
parameters are listed in Tab. \ref{tab:max} and \ref{tab:min}.

In the colour-magnitude relation we confirm the weak intrinsic
blueing with increasing redshift from $z \sim 0.4$ to $z\sim 2$
already found by G05 for both populations although formally only at
2$\sigma$ level. In the redshift bins where the statistics is poor,
the minimum is extrapolated from the other redshift bins. From the
bins $z=0.4-0.7$ and $z=0.7-1.0$ we extrapolate the minimum value
$<U-V>_{20}=1.58$ in the redshift interval $z=0.2-0.4$, and from the
bins $z=0.7-1.0$ and $z=1.0-2.0$ we extrapolate $<U-V>_{20}=1.23$ in
the interval $z=2-3.5$.

We found no appreciable redshift evolution in the SSFR distribution,
so in order to increase statistics we have performed the fit in the
larger redshift intervals $z=0.2-1$ and $1-3.5$. Moreover, there is
no appreciable dependence of SSFR on the absolute magnitude, so the
colour-magnitude relation is not related to similar trends in the
specific star-formation rate.

\subsection{Discussion on the selection
criteria}\label{sec:bimodality2}

One notes in the colour/SSFR-magnitude relations the presence of a
conspicuous number of intrinsically faint galaxies with relatively
red colours. They are red with respect to the locus of separation of
the two populations although, because of the colour-magnitude
relation, their colours are typical of the bright ($M_B\sim -22$)
blue galaxies. In terms of SSFR these galaxies show intermediate
values between star forming and early type galaxies. The presence of
a large number of galaxies belonging to this intermediate population
dominates the shape of the LF of the red/early type galaxies at the
faint end, as shown in the next sections.

\begin{figure}[htb]
\resizebox{\hsize}{!}{\includegraphics{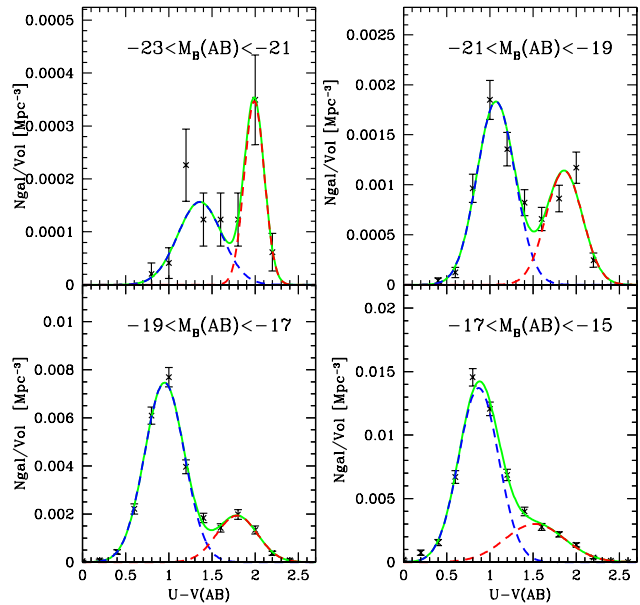}} \caption{Volume
corrected colour histogram, in four bins of magnitude, in the
redshift interval $0.4<z<0.7$. The continuous lines are the
double-gaussian fit to the colour distributions.}\label{fig:bigauss}
\end{figure}

\begin{figure}[htb]
\resizebox{\hsize}{!}{\includegraphics{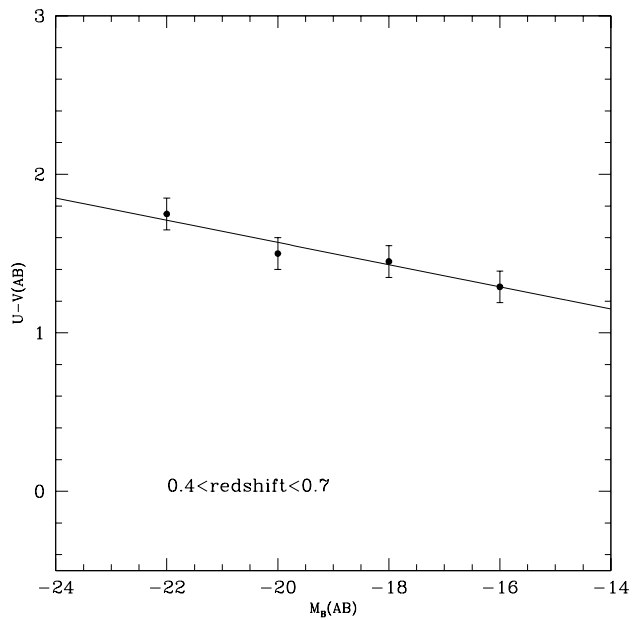}} \caption{Points are
the minima obtain by the fits shown in Fig \ref{fig:bigauss}.
Continuous line is the linear fit to these minima.
}\label{fig:retta}
\end{figure}

However, since the colour dispersion of the blue sequence broadens
at faint magnitudes, the assumption of a linear parameterisation for
the minimum could imply a bias for the selection of the red sample
\citep[for a detailed analysis of the blue sequence properties
see][]{labbe2007}. For this reason we have also performed an
analysis deriving the minimum without any assumption on its
dependence on the rest-frame luminosity. We have concentrated our
analysis in the redshift interval 0.4-0.7 which have sufficient
statistics at faint magnitudes. We have obtained a volume corrected
colour distribution in four magnitude bins, and for each colour
distribution we have fit a double-gaussian function as shown in Fig.
\ref{fig:bigauss}. We have adopted the intersection of the two
gaussians as minimum. In this way we have verified that the locus of
the minimum is well described by a linear behaviour: performing a
linear fit to these points we have found the following relation
$<U-V>=-0.07 \cdot (M_B+20)+ 1.57$ (see Fig. \ref{fig:retta}). This
relation is very similar to our standard analysis and does not
produce a substantial variation of our results (see also Section
\ref{subsec:redblue} and Fig. \ref{fig:lfred0406}). As described
above, this analysis can not exclude a contamination from the
star-burst galaxies reddened by dust, for this reason we have also
performed an analysis of the LF on a homogenous sample of galaxies
selected using SSFR distribution.


\begin{table}
\caption{Parameters of the relation between the loci of the maxima
and the absolute B magnitude.}\label{tab:max} \centering
\begin{tabular}{lr@{$\pm$}rr@{$\pm$}rr@{$\pm$}r}
\hline \hline \noalign{\smallskip} \multicolumn{1}{c}{$\langle z
\rangle$}&\multicolumn{2}{c}{\A}
&\multicolumn{2}{c}{m$_{red/early}$} &
\multicolumn{2}{c}{m$_{blue/late}$}\\
\noalign{\smallskip} \hline \noalign{\smallskip}
\multicolumn{7}{c}{U-V} \\
\noalign{\smallskip} \hline \noalign{\smallskip}
0.4 0.7&$ -0.08 $ & 0.01   & 1.79  & 0.05  & 1.09 & 0.03 \\
0.7 1.1&$ -0.08 $ & 0.01   & 1.74  & 0.05  & 0.93 & 0.03  \\
1.1 2.0& \multicolumn{2}{c}{$ -0.08$}   & 1.69  & 0.05 & 0.90 & 0.03   \\
\noalign{\smallskip} \hline \noalign{\smallskip}
\multicolumn{7}{c}{SSFR} \\
\noalign{\smallskip} \hline \noalign{\smallskip}
0.2 1.1&$ 0.019 $ & 0.03  & -11.59 & 0.20 & -9. & 0.03    \\
1.1 3.5& \multicolumn{2}{c}{$0.019$} & -11.55  & 0.20 & -9.07  & 0.03      \\
\noalign{\smallskip} \hline \noalign{\smallskip}
\end{tabular}
\end{table}

\begin{table}[htb]
\caption{Parameters of the relation between the locus of the minimum
and the absolute B magnitude.} \label{tab:min} \centering
\begin{tabular}{lr@{$\pm$}rr@{$\pm$}r}
\hline \hline $\langle z \rangle$ & \multicolumn{2}{c}{\A} &
\multicolumn{2}{c}{minimum}\\
\noalign{\smallskip} \hline \noalign{\smallskip}
\multicolumn{4}{c}{U-V} \\
\noalign{\smallskip} \hline \noalign{\smallskip}
0.2 0.4& \multicolumn{2}{c}{$-0.08$ }  & 1.58  & 0.10 $^*$\\
0.4 0.7& $-0.08$& 0.01     & 1.51  & 0.07  \\
0.7 1.1& $-0.08$ & 0.01    & 1.43  & 0.06   \\
1.1 2.0& \multicolumn{2}{c}{$-0.08$ } & 1.36  & 0.07   \\
2.0 3.5& \multicolumn{2}{c}{$-0.08$ } & 1.23  & 0.10  $^*$\\
\noalign{\smallskip} \hline \noalign{\smallskip}
\multicolumn{4}{c}{SSFR} \\
\noalign{\smallskip} \hline \noalign{\smallskip}
0.2 1.1& 0.019 & 0.020  & -10.41  & 0.20\\
1.1 3.5& \multicolumn{2}{c}{$0.019$} & -10.43  & 0.20   \\
\noalign{\smallskip} \hline \noalign{\smallskip}
\end{tabular}
\\ $^*$  Extrapolated value
\end{table}

\section{LUMINOSITY FUNCTION}

\subsection{The  Statistical Analysis}\label{subsec:STATISTIC}

The LF is computed with the procedure described in G05. We have
applied to our sample an extended version of the standard
$1/V_{max}$ algorithm \citep{avni1980}. As in the previous paper, we
have used a combination of data derived from regions in the field
with different magnitude limits. Indeed, for each object and for
each $j$-th region under analysis a set of effective volumes
$V_{max}(j)$ is computed. For a given redshift interval $(z_1,z_2)$,
these volumes are enclosed between $z_1$ and $z_{up}(j)$, the latter
being defined as the minimum between $z_2$ and the maximum redshift
at which the object could have been observed within the magnitude
limit of the $j$-th field. The galaxy number density $\phi(M,z)$ in
each $(\Delta z,\Delta M)$ bin can then be obtained as:
\begin{equation}
\phi(M,z)=
\frac{1}{\Delta M}\sum_{i=1}^{n}\left[ \sum_j
\omega(j)\int_{z_{1}}^{z_{up}(i,j)} \frac{dV}{dz}dz \right]^{-1}
\end{equation}
where $\omega(j)$ is the area in units of steradians corresponding
to the field $j$, $n$  is the number of objects in the chosen bin
and $dV/dz$  is the comoving volume element per steradian.

The Poisson error in each LF magnitude bin was computed adopting the
recipe by \cite{Gehrels1986}, valid also for small numbers. The
uncertainties in the LF due to the photometric uncertainties and to
the degeneracy of the spectral models used to derive redshifts were
computed with the same Monte Carlo analysis described in the
previous section. The uncertainties obtained and the Poisson errors
have been added in quadrature.

The $1/V_{max}$ estimator for the LF can in principle be affected by
small scale galaxy clustering. For this reason a parametric maximum
likelihood estimator is also adopted which is known to be less
biased respect to small scale clustering \citep[see][]{heyl1997}.

The parametric analysis of the galaxy LF has been obtained from the
maximum likelihood analysis assuming for different galaxy
populations different parameterisation $\phi$ for the LF. The
maximum likelihood method used here represents an extension of the
standard \cite{sandage1979} method where several samples can be
jointly analysed and where the LF is allowed to vary with redshift.
A more detailed description can be found in G05 and a formal
derivation of the maximum likelihood equation is shown in
\cite{heyl1997}.

To describe the B-band LF of the total sample and that of blue/late
galaxies we assume a Schechter parameterisation:
$$
\begin{array}{c}
\phi(M,z)=0.4 \cdot \ln(10) \phi^*
[10^{0.4(M^*-M)}]^{1+\alpha}\cdot \\
\exp[-10^{0.4(M^*-M)}]
\end{array}
$$
As previous analysis have shown that the evolution of the global
galaxy LF is manly driven by luminosity evolution, we have adopted
an $M^*$ that is redshift dependent \citep[see
aslo][]{heyl1997,giallo2005,gabasch2006}. We have adopted the
parameterisation that better matches our data in this range of
redshifts:
\begin{equation}\label{eq:mag}
M^*(z)=\left\{
\begin{array}{ll}
M_0-M_1\cdot z \ \mbox{if} \ z\le z_{cut} \\
M_0-M_1\cdot z_{cut}\ \mbox{if} \ z > z_{cut}
\end{array} \right.
\end{equation}
The slope $\alpha$ is kept constant with redshift since it is well
constrained only at low-intermediate redshift ($z<1$). For all the
sample here considered, we checked that the density parameter $\phi$
had no significative variation if it had been evolved with the
parametric form described in G05, thus we keep it as constant.

As we will see in Fig. \ref{fig:lfred1} and \ref{fig:lfred2} this is
not a good description for the red/early population for which we
assume a double Schechter form, as frequently adopted in similar
cases in literature \citep[e.g.][]{popesso}:
\begin{equation}\label{eq:sch}
\begin{array}{c}
\phi(M)=\phi_f(M)+\phi_b(M)=\\
\phi^* \cdot 0.4 \cdot \ln(10)\cdot\\
([10^{0.4(M^*_{f}-M)}]^{1+\alpha_{f}}\exp[-10^{0.4(M^*_{f}-M)}]+\\
([10^{0.4(M^*_{b}-M)}]^{1+\alpha_{b}}\exp[-10^{0.4(M^*_{b}-M)}])
\end{array}
\end{equation}
For a quantitative evaluation of the density evolution at the bright
end, we have evolved the bright Schechter with a redshift dependent
normalisation:
\begin{equation}\label{eq:schevol}
\phi(M,z)=\phi_f(M) +norm(z) \cdot \phi_b(M)
\end{equation}
where
\begin{equation}\label{eq:norm}
norm(z)=\left\{
\begin{array}{ll}
1 \  \mbox{if} \ z<z_{cut} \\
 z^{\gamma} \cdot (z_{cut})^{-\gamma} \ \mbox{if} \ z \ge z_{cut}
\end{array} \right.
\end{equation}

We used the MINUIT package of the CERN library \citep{james1995} for
the minimisation. The errors in Tab. \ref{tab:evolblueall} and
\ref{tab:redlf} are calculated for each parameter, independently of
the others.

\subsection{B-band luminosity  function from  $z\sim 0.2$  to $z  \sim
3.5$ and the comparison with the other surveys}

\begin{figure}
\resizebox{\hsize}{!}{\includegraphics{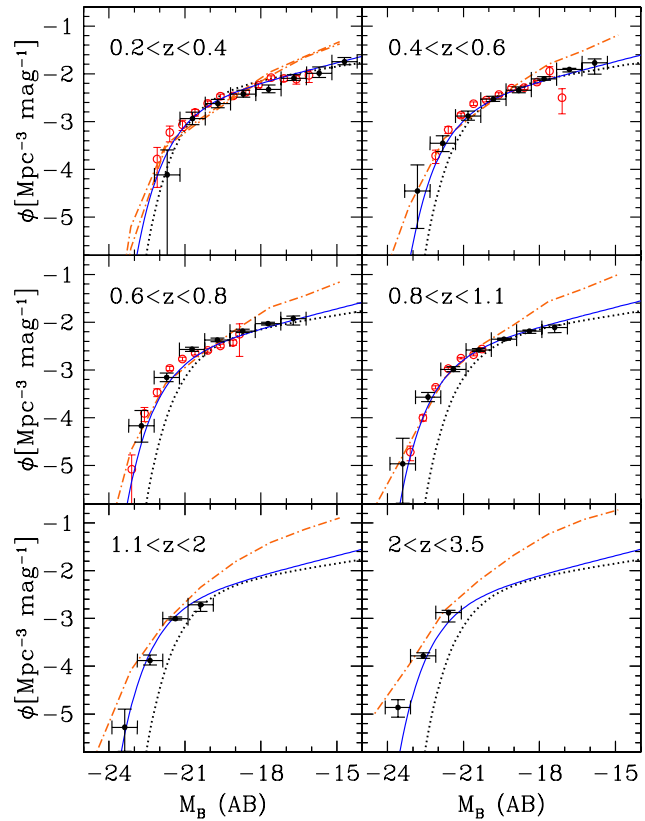}} \caption{Total LF
as a function of redshift. The continuous curves come from our
maximum likelihood analysis. The  dotted line is the local LF
\citep{norberg2002}. The filled circles are the points obtained with
$1/V_{MAX}$ method. The empty circles come from the DEEP2 survey
\citep{willmer}. The results from the COMBO17 and VDDS surveys are
consistent with the DEEP2 results and are omitted in the plot. The
dashed-point line is the model of \cite{menci2006}.}\label{fig:ftot}
\end{figure}

The evolution of the total LF is shown in Fig. \ref{fig:ftot}. To
compare our high $z$ results and our fits with local values we also
show the local LF from Two-Degree Field Galaxy Redshift Survey
\citep[2dFGRS; ][dotted line in Fig. \ref{fig:ftot}]{norberg2002}.

The $1/V_{max}$ analysis shows that the main kind of evolution is
due to a brightening of the LF with redshift. We have applied the ML
analysis to the sample using the evolutionary form of the Schechter
LF described in eq. \ref{eq:mag} where pure luminosity evolution is
allowed up to a maximum redshift beyond which the LF keeps constant.
The results in Tab. \ref{tab:evolblueall} imply that the LF is
subject to a mild luminosity evolution only up to $z\sim 1$ ($\Delta
M^*\sim 0.7$ in the $z=0.2-1$ interval). At higher $z$ the LF
appears constant with redshift although at $z\sim 3$, in the
brightest bin, a slight excess is present. In any case, the adopted
evolutionary model is acceptable at 2$\sigma$ level using the
standard $\chi^2$ test.

We also show a comparison with the DEEP2 Redshift Survey
\citep[D2RS,][empty points in Fig.
\ref{fig:ftot}]{davis2003,willmer}. Although the comparison of the
LFs was performed on data taken from surveys having different
magnitude limits and redshift estimates (photometric and
spectroscopic), the agreement is very good in the overlapping
magnitude regions and in all the redshift bins. A general good
agreement is also found with the LF derived from the VVDS survey by
\cite{ilbert}, although from our ML analysis we do not have any
evidence of steepening of the faint end slope up to $z\sim 2$, as
suggested by them. We have compared our results with other
photometric redshift surveys like the COMBO-17 survey by
\cite{bell2004} \citep[see also][]{faber}. A good agreement is found
in the overall redshift interval and in the appropriate magnitude
interval. The comparison with the FORS Deep Field (FDF)
\citep{gabasch2004} is less straightforward, because of the
different redshift intervals used. An acceptable agreement is found
up to $z\sim 1$. At higher redshifts the FDF luminosity function
shows an excess of very bright objects with respect to our values as
well as the COMBO17 and DEEP2 LFs.

In Fig. \ref{fig:ftot} we included the LFs derived from our
hierarchical model described in sec. \ref{sec:model} \citep[see also
][]{menci2005,menci2006}. The effect of the UV background is
effective in flattening the predicted shape of the LF at the faint
magnitudes providing an agreement better than that obtained before
\citep{poli2003}.

\begin{table*}[htb]
\caption{Luminosity Function Parameters for the total and blue/late
type sample}\label{tab:evolblueall} \centering
\begin{tabular}{clr@{$\pm$}rr@{$\pm$}lr@{$\pm$}lr@{$\pm$}lr@{$\pm$}lc}
\hline\hline \noalign{\smallskip}
\multicolumn{1}{c}{Type}&\multicolumn{1}{c}{$z$} &
\multicolumn{2}{c}{$M_0^*$} & \multicolumn{2}{c}{$\alpha$} &
\multicolumn{2}{c}{$M^*_1$} &\multicolumn{2}{c}{$z_{cut}$}&
\multicolumn{2}{c}{$\phi_0^*$} &\multicolumn{1}{c}{$N$}\\
\noalign{\smallskip} \hline \noalign{\smallskip}
all &0.2-3.5     & $-20.54$ & $0.12 $ & $-1.34$ & $0.01 $ &$ 0.93 $ &$0.13$ &0.96 &0.02&\multicolumn{2}{c}{0.0031}& 5115 \\
\noalign{\smallskip} \hline \noalign{\smallskip}
blue&0.2-3.5     & $-19.97$ & $0.20 $ & $-1.39$ & $0.02 $ &$ 1.43 $ &$0.27$ &0.98 &0.05&\multicolumn{2}{c}{0.0024} &4127\\
late&0.2-3.5     & $-20.06$ & $0.12 $ & $-1.36$ & $0.02 $ &$ 1.29 $ &$0.12$ &0.99 &0.02&\multicolumn{2}{c}{0.0029} &4612\\
\noalign{\smallskip} \hline \noalign{\smallskip}
\end{tabular}
\end{table*}
\subsection{Luminosity function for the blue/late and red/early
galaxies}\label{subsec:redblue}

In this section we show the shapes and evolutionary behaviours of
the luminosity functions derived for the blue/late and red/early
galaxy populations respectively. We have adopted the empirical
colour/SSFR selection described in section \ref{sec:bimodality} to
separate the two populations.

\begin{figure}[htb]
\resizebox{\hsize}{!}{\includegraphics{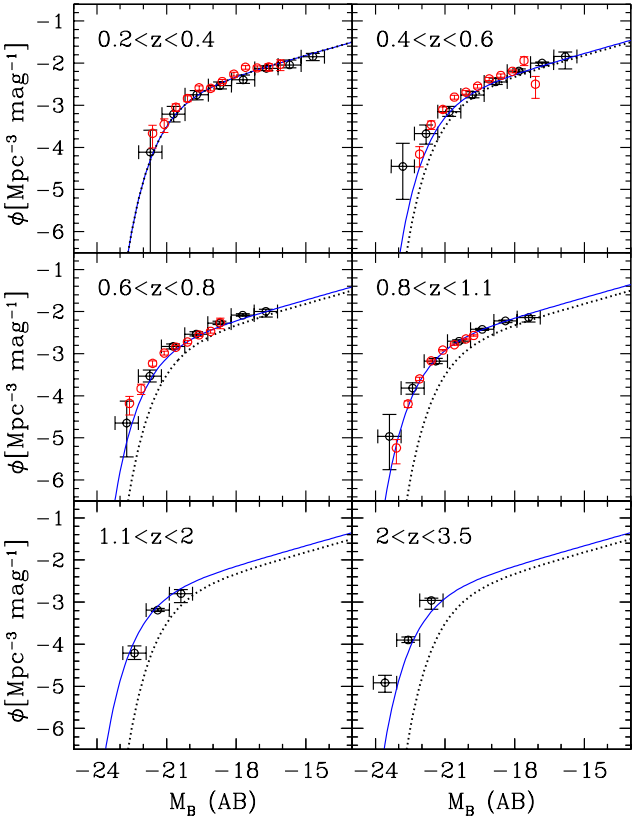}} \caption{LFs of the
blue galaxies as function of redshift. The continuous curves comes
from our maximum likelihood analysis. The dotted line is our fit at
$z\sim 0.3$, reported for comparison in all the redshift bins. The
big filled circles are the points obtained with $1/V_{MAX}$ method.
The little empty circle are points from the LFs by
\cite{willmer}.}\label{fig:lfblu}
\end{figure}

The shape and redshift evolution of the blue LF is shown in Fig.
\ref{fig:lfblu} where both the $1/V_{max}$ data points and the
curves derived from the ML analysis are represented. The best fit
parameters together with their uncertainties are shown in Tab.
\ref{tab:evolblueall}. The LFs of the late populations are very
similar to the blue ones and we do not show the figure.

As for the total sample, we found that the blue population is well
represented by the same type of luminosity evolution although
faster, with $\Delta M^*\sim 1.15$ in the $z=0.2-1$ interval. The
faint end slope appears steeper. This is due of course to the fact
that the blue population dominates the volume density of the total
sample at any redshift.

We first compare  our results with those of the spectroscopic survey
D2RS \citep{faber,willmer}. We note that the colour selection they
use to separate the two populations is based on a $U-B$ vs $M_V$
colour-magnitude relation. We have verified in our sample that their
colour selection is very similar to our criterion. In fact, if we
adopt their selection on our sample, we reproduce almost the same
blue/red galaxy subsamples obtained with our cut. Their LFs are in
good agrement with our results as shown in Fig. \ref{fig:lfblu}
(data with error bars). We then compare our results to those of
COMBO-17. They use as a selection criterion the colour $U-V$ vs
$M_V$, which nearly corresponds to that used by \cite{willmer}. The
agreement with our results is good.

A direct comparison with the blue/red LFs by \cite{marchesini2006} in the redshift interval $2<z<3.5$ is not possible since
they use a colour separation bluer than our criterion by 0.2 mag
providing a LF with a lower normalisation.

The LF predicted by our hierarchical model was not included in Fig.
\ref{fig:lfblu} since it is not appreciably different from that of
the total sample.

\begin{figure}[htb]
\resizebox{\hsize}{!}{\includegraphics{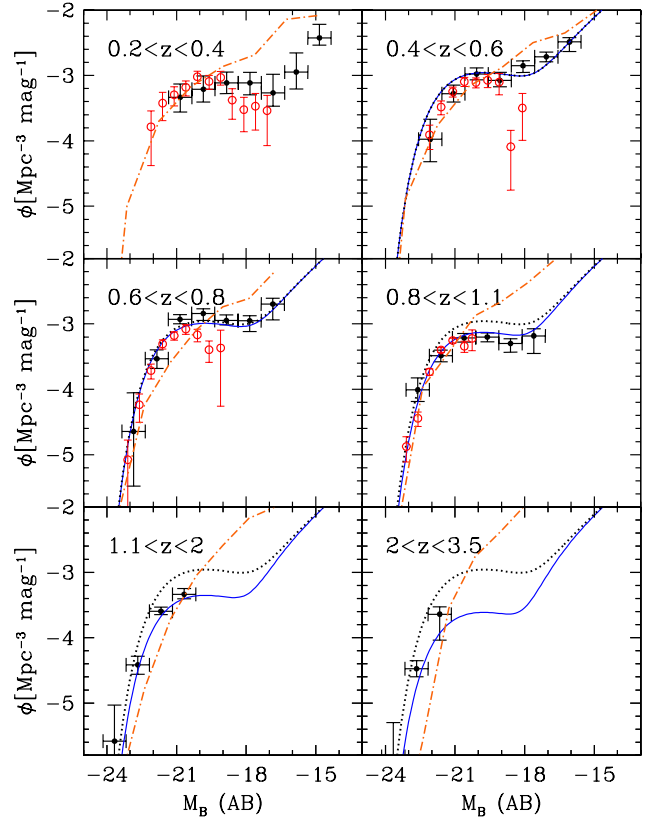}} \caption{LFs of the
red galaxies as function of redshift. The continuous curves comes
from our maximum likelihood analysis, the first bin of redshift,
which have to low statistic, has been excluded from this evolutive
analysis. The dotted line is our fit at $z\sim 0.5$, reported for
comparison in all the redshift bins. The big filled circles are the
points obtained with $1/V_{MAX}$ method. The little empty circle are
points from the LFs by \cite{willmer}. The dashed-point orange line
is the model of \cite{menci2006}}\label{fig:lfred1}
\end{figure}

\begin{figure}[htb]
\resizebox{\hsize}{!}{\includegraphics{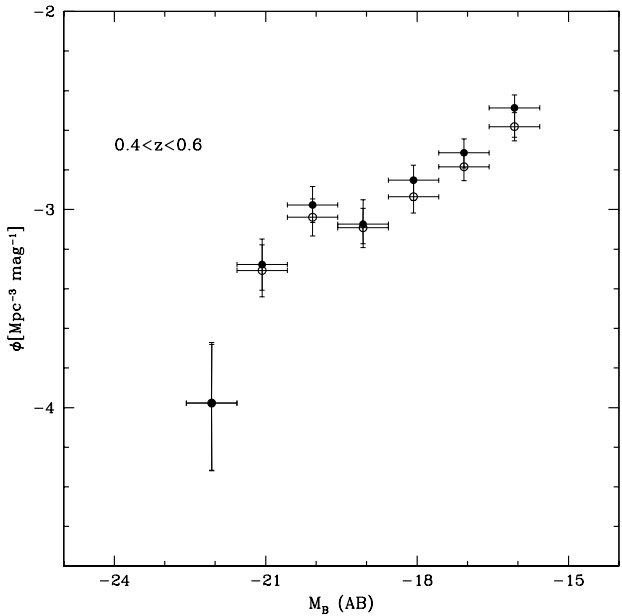}} \caption{LFs of the
red galaxies. With the filled circle is indicated the LF obtained
with the linear color selection in Tab. \ref{tab:min}. The empty
circles indicate the LF obtained from the color histograms in bins
of magnitude (see Fig. \ref{fig:bigauss} and Fig. \ref{fig:retta}).
}\label{fig:lfred0406}
\end{figure}

\begin{figure}[htb]
\resizebox{\hsize}{!}{\includegraphics{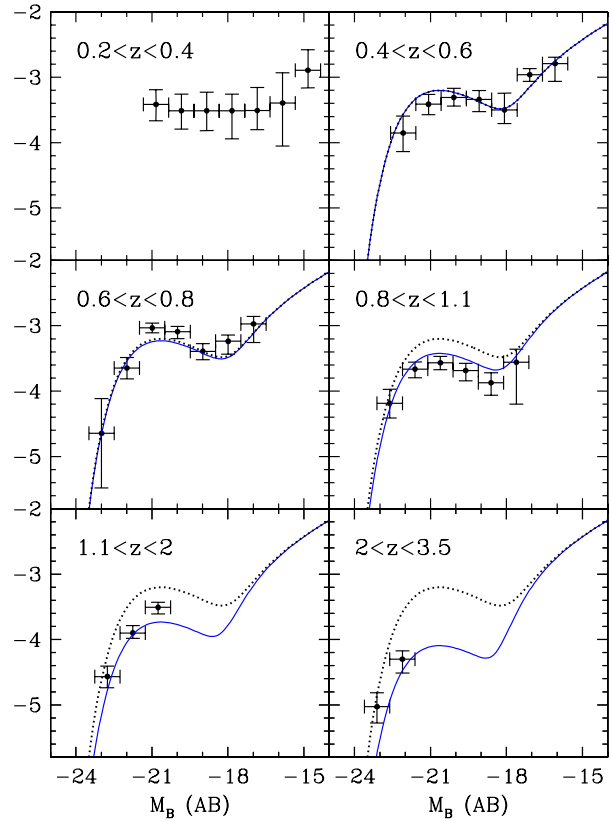}} \caption{LFs of the
early type galaxies as function of redshift. The continuous curves
comes from our maximum likelihood analysis. The dotted line is our
fit at $z\sim 0.5$, reported for comparison in all the redshift
bins. The big filled circles are the points obtained with
$1/V_{MAX}$ method.}\label{fig:lfred2}
\end{figure}

Concerning the red/early populations our GOODS-MUSIC sample allows a
sampling of the red LF down to $M\sim -16$ up to $z\sim 0.8$
providing for the first time a direct evaluation of the faint shape
of the LF. This is due to the deeper $K$ magnitude limit respect to
that used in G05, allowing the evaluation of the rest frame $U-V$
colour at magnitudes as faint as $M_B\sim -16$. At variance with
previous works which involve shallower samples, a peculiar LF shape
is present at $z<0.8$ with a minimum and a clear upturn at $M_B \geq
-18$ (Fig. \ref{fig:lfred1}).

This overabundance of faint objects with respect to the
extrapolation of the Schechter function was already found in the LF
of local early type galaxies derived from the 2dF survey
\citep{madgwick2002}. Although they used a different separation
criterion, their subsample called type 1 is not so different from
our subsample, being equivalent to a morphological sample of E, S0
and Sa. A similar upturn was also found in the local LF of red
galaxies derived from the Sloan survey by \cite{blanton2005}.

We have checked that the characteristic shape found is not
critically dependent on the specific choice of the colour-magnitude
or SSFR -magnitude separation. Indeed, changing the parameters of
the linear fit in colour-SSFR separation at $2-\sigma$ level, the
shape and, in particular, the upturn do not change appreciably. We
have also found that the overall shape of the LF does not change, if
we use a separation between blue and red galaxies found with an
analysis without any assumption about the parametric dependence on
the the rest-frame magnitude (see section \ref{sec:bimodality2}).
Moreover, fitting with a double gaussian function the bimodal
distribution in the faintest magnitude interval (see Fig.
\ref{fig:bigauss}, $-17<MB<-15$) the contamination by any blue
population in the locus of the red population is not so strong. In
other words the FWHM of the blue gaussian distribution is relatively
narrow. In this case the expected fraction of blue galaxies expected
redward of the selected minimum is only 14\% of the red population
in the same colour region. For this reason the shape of the red LF
shown in Fig.8 relative to the redshift interval $0.4<z<0.6$ remains
almost unchanged in the two fainter magnitude bins if we remove this
small fraction.

Also for the red population we have compared our LFs with that
derived from the major surveys of colour selected galaxies. In Fig.
\ref{fig:lfred1} we show the LF from the spectroscopic survey of
\cite{willmer}. For the red galaxies the agreement is good for
$M_B<-20$, where the incompleteness is negligible in their sample
\cite[see fig 8, ][]{willmer}. Their shallower sample can not probe
the raise at the faint end present in our deeper data. The same
holds for the two photometric surveys of \cite{bell2004} and
\cite{brown2006}.

Concerning the parametric analysis of the evolutionary LFs of the
red/early galaxy population, given the excess of faint objects a
single Schechter shape does not provide an acceptable description of
the data. For this reason we have adopted a double Schechter
function as described in eq. \ref{eq:schevol}. The best fit
parameters are shown in Tab. \ref{tab:redlf}. The best fit value of
the brighter Schechter slope is rather flat ($\alpha_b \sim -0.7$)
in agreement with what found in G05 and in the Bell et al. sample.
The fainter slope is steeper approaching the value $\alpha_f\sim
-1.8$. As for the redshift evolution we have adopted the density
evolution law described in eq. \ref{eq:norm} where the Schechter
shape at the faint end is kept constant at all redshifts. The
brighter one is constant only up to a given redshift $z_{cut}$
beyond which decreases as a power law in redshift. We find a
constant density up to $z\sim 0.7$ and thereafter a decrease by a
factor $\sim 5$ up to $z\sim 3.5$.

The LF evolution of the early galaxies selected from their SSFR
value is very similar to that of the red ones although the high
redshift density evolution is more pronounced with a decrease by a
factor $\sim 10$ in the interval $z=0.7-3.5$. This difference is
caused by the presence, in the red sample at higher redshift, of a
high fraction of galaxies having SEDs consistent with those of a
dusty and starburst galaxy. This fraction amounts to $\sim 70$\% at
$M_B\sim-21.5$ and $z\sim 3$.


\begin{table}
\centering \caption{Luminosity Function Parameters for red and early
type galaxies, in the redshift interval 0.4-3.5}\label{tab:redlf}
\begin{tabular}{cr@{$\pm$}lr@{$\pm$}l}
\hline\hline \noalign{\smallskip}
\multicolumn{1}{c}{parameter}&\multicolumn{2}{c}{Red} &\multicolumn{2}{c}{Early}         \\
\noalign{\smallskip} \hline \noalign{\smallskip}
$N       $   &\multicolumn{2}{c}{913}      &\multicolumn{2}{c}{472}              \\
$\phi_0^*$   &\multicolumn{2}{c}{0.0020}    &\multicolumn{2}{c}{0.0017}              \\
$\gamma  $   &-1.045   &0.13                &   -1.45 & 0.16              \\
$z_{cut} $   & 0.65    & 0.01               & 0.67   & 0.01                       \\
$\alpha_b$   & -0.76   & 0.06               & -0.40&0.1                      \\
$M_b^*   $   & -21.29  & 0.1                & -21.22 & 0.12                         \\
$\alpha_f$   &  -1.77  & 0.2                &  -1.54  & 0.22                                  \\
$M_f^*   $   & -17.04  & 0.12               & -17.06 & 0.12                          \\
\noalign{\smallskip} \hline \noalign{\smallskip}
\end{tabular}
\end{table}
Finally, the comparison with our hierarchical CDM model shows a
slight flattening of the LF at intermediate luminosities. This is because the
red population is mainly contributed by galaxies with larger M/L
ratios; these are mainly attained in massive objects (due to the
ineffectiveness of gas cooling, to their earlier conversion of gas
into stars, and to the effect of AGN feedback) or in low-mass
objects (due to the gas depletion originated from the different
feedback mechanisms, particulary effective in shallow potential
wells).
However, the model still overpredicts the LF at faint
magnitudes; we shall investigate the origin of such an effect
(probably originated at high-redshifts) in a future paper.

\subsection{Luminosity densities}

\begin{figure}
\resizebox{\hsize}{!}{\includegraphics{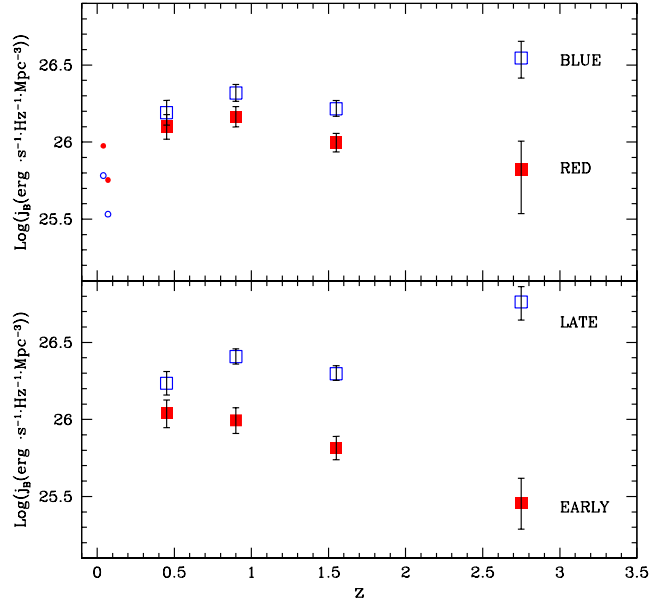}} \caption{Upper
panel: Galaxy luminosity density for red (filled square) and blue
galaxies (empty square). Lower panel: the same of the upper panel
but for early type galaxies (filled square) and late type galaxies
(empty square). Small points are the local luminosity density
estimates by \cite{madgwick2002} (z=0.04) and \cite{bell2003}
(z=0.07). }\label{fig:numbuv}
\end{figure}

To compare in a global way the redshift evolution of the blue and
red galaxies we have computed the B band luminosity densities of the
two populations as a function of redshift. To make  the comparison
homogeneous we have computed the contribution of the same bright
population with $M_B(AB)\leq -20.2$ at all redshifts. The results
are shown in Fig. \ref{fig:numbuv} for the blue/red and late/early
populations. The redshift bins are selected to have a comparable
number of objects in the considered magnitude range. In fact for the
highest redshift bin the lowest luminosity of the data is
$M_B(AB)=-21$. For this reason we have added the contribution of the
fainter sources using the extrapolation of the parametric LF.

The uncertainty associated with the luminosity density is the sum in
quadrature of the error estimated through jackknife\footnote{The
jackknife analysis is performed recomputing the statistic estimate
leaving out one observation at a time from the
sample.}\citep{efron1982} analysis and of the error obtained from
the MonteCarlo analysis as described in sec. \ref{subsec:STATISTIC}.
The first contribution is associated with the clustering properties
of the field and the second with the photometric uncertainties.
Local luminosity densities, obtained from the integration of the LFs
by \cite{madgwick2002} (z=0.04) and \cite{bell2003} (z=0.07), are
also included for comparison.

The blue/late galaxies increase steadily their luminosity density up
to $z\sim 3.5$ while the luminosity density of the red/early
population is nearly constant up to $z\sim 1$ and then decreases by
a factor $\sim 3$ at $z\sim 3.5$. This confirms our previous result
presented in G05, which shows an appreciable decline of the bright
red/early population only at $z>1$. As already noted this is not in
qualitative contrast with the hierarchical scenario, since the
bright and hence massive early population is developing early in the
cosmic time in specific overdense regions where the evolution is
also accelerated by merging processes. However, the detailed
quantitative agreement of the LFs predicted by specific models
depends on the details of the main physical processes and a
satisfactory agreement is not yet obtained especially at the fainter
magnitudes (see previous section).

\section{Red/early faint galaxies}
\begin{figure}
\resizebox{\hsize}{!}{\includegraphics{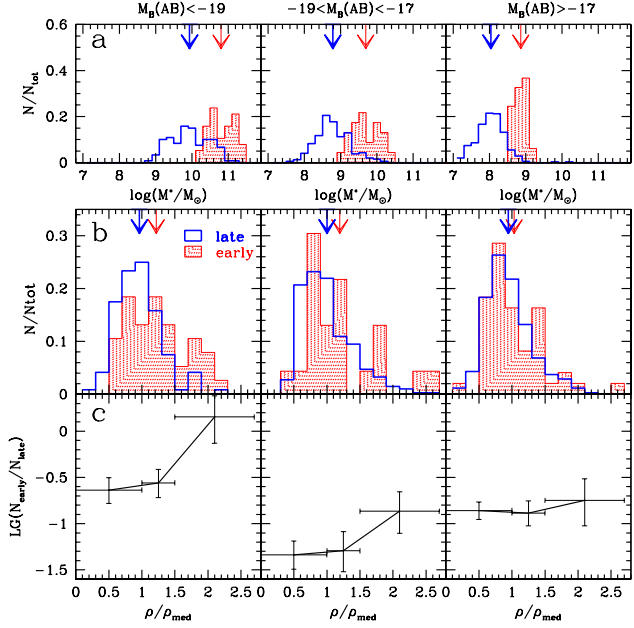}} \caption{{\bf
panel a}: Stellar mass distribution for the early and late
population, the area of each histogram is normalised to unity. The
arrows indicate the mean value of the stellar mass for the early
population (thin arrow) and for the late population (tick arrow)
{\bf panel b}: as the panel a but for the distribution of
$\rho/\rho_{med}$ {\bf panel c}: Early type galaxies vs. late type
galaxies fraction as function of the density contrast
($\rho/\rho_{med}$).}\label{fig:density}
\end{figure}

As shown in the previous section, the presence of an upturn in the
LF of the red galaxy population at faint magnitudes is a new feature
emerging from the analysis of deep NIR selected galaxies, with
respect to previous works at this redshift. To derive hints on the
nature of the population responsible for this excess we have
analysed their colour and spatial properties.

The peculiar shape of the LF represented by a double Schechter form
appears similar to that obtained for galaxies in local rich
clusters. Indeed recent studies of the total galaxy luminosity
functions in clusters selected from the RASS-SDSS survey show an
upturn at faint $M_r$ which depends on the distance from the cluster
center \citep{popesso}. The main contribution to this local excess
comes from the red population selected with $u -r>2.22$
\citep{strateva2001}. They also noted that the ratio between red
and blue galaxies increases with the density in the clusters.

It is interesting to explore whether a dependence on the environment
is present for our faint early subsample. To this end we have
adopted a 2-D density analysis of the 20 first-neighbour method. We
describe here briefly the procedure adopted to derive densities,
while we refer to \cite{trevese2006} for a detailed description of
the method. A 2-D density has been assigned to each object and a
density map has been derived in the field in the redshift interval
$z=0.4 -0.6$.

A surface density
\begin{equation}
 \Sigma_n = n/(\pi D_{p,n}^2)
\end{equation}
of galaxies was computed considering the projected distances
$D_{p,n}$ to the n-th nearest neighbour, as done by Dressler
\citep{dress1,dress2}. We divide the survey area in cells whose
extension depends on the observational accuracy. For each cell we
count neighbouring objects at increasing radial distance until a
number $n=20$ of objects is reached. In counting galaxies we must
take into account the increase of limiting luminosity with
increasing redshift for a given flux limit. If $m_{lim}$ is the
limiting (apparent) magnitude in a fixed observing band, at each
redshift $z$ we detect only objects brighter than an absolute
magnitude $M_{lim}(z)$, decreasing (brightening) with $z$ . We
assume the lower limit of the interval as reference redshift $z_c$
below which we detect all objects brighter than the relevant $M_c
\equiv M_{lim}(z_c)$, which is the magnitude of the faintest galaxy
in the B band among the objects we are considering. At $z > z_c$ the
fraction of detected objects is:
\begin{equation}
s(z)=\frac{\int^{M_{lim(z)}}_{-\infty}\Phi(M)dM}{\int^{M_c}_{-\infty}\Phi(M)dM}
\end{equation}
where $\Phi(M)$ is the type and z-dependent galaxy B-band luminosity
function presented above. Thus, in evaluating the galaxies number
density, we apply a {\it limiting magnitude correction} by assigning
a weight $w(z)=1/s(z)$ to each detected galaxy of redshift $z$. In
this way we have evaluated the surface density field and assigned a
density value to each object.

For our analysis we have selected galaxies in three magnitude
regions; an intermediate region, $ -19<M_B(AB)<-17$, where the LF of
the early population is flat, and two external steeper regions at
the bright and faint end of the LF, $M_B(AB)<-19$ and $M_B(AB)>-17$
respectively.

First of all  we note that  early galaxies represent the most massive
galaxies in each luminosity interval (Fig. \ref{fig:density} panel a).
In particular even at the faint end of the LF the early population  is
clearly segregated in stellar mass with values one order of magnitude  greater
on average with respect to  the late  population.

Looking to the brightest fraction, a clear difference as a function of
the density of the environment is found between the early and the late
populations. In particular early galaxies tend to populate regions  of
higher  density.  This   is  shown  in   panels  b,c  where   the  two
distributions are  represented as  a function  of the  density of  the
environment. The  ratio early/late  increases with  density since  the
average  density  of  the early  galaxies  is  somewhat greater  (1.4)
with respect to the one of the late population.

This  different  behaviour   becomes  less  evident   with  decreasing
luminosity and almost disappears at the  faint end of the two LFs.  We
note in this respect that the limited area covered by our sample  does
not allow an evaluation of  the environment dependence up to  the high
densities typical  of clusters, like those  probed by  e.g. the  Sloan
survey.

Thus, the scenario that emerges is one where major evolutionary
differences between the early and late populations act in the
relatively bright galaxies with $M<-17$ producing the largest
differences in the shapes of the two LFs in the interval
$-21<M_B(AB)<-18$ (flat shape for the early, steep for the late). At
faint magnitudes the two LFs tend to converge to the same volume
density. From this analysis the characteristic shape of the
red/early LF does not seem to depend strongly on the environmental
properties.

\section{Summary}

We have used a composite sample of galaxies selected in deep NIR
images, obtained from the GOODS public survey, to study the evolution
of the galaxy LFs of red/early and blue/late galaxy populations
selected using the colour and SSFR statistical properties of the sample.

The  observed  $U-V$  colour  and  SSFR  distributions  show  a
clear bimodality up to $z\sim 3$, confirming the results obtained in
G05  at a  higher  level of  statistical  confidence. We  found  a
trend  with redshift  for  the  colour magnitude  distribution  with
an intrinsic blueing of about 0.15 mag.  in the redshift interval
$z=0.4-2.0$  for both populations. This observed bimodality can be
explained  in  a  hierarchical clustering scenario  as  due to the
different  star formation histories of the red/early  and blue/late
galaxies, see e.g. \cite{menci2006}.

For the total and the blue/late sample the LF is well described by a
Schechter function and shows a mild luminosity evolution in the
redshift interval $z=0.2-1$ (e.g. $\Delta M^* \sim 0.7$ for the
total sample; $\Delta M^*\sim 1$ for the blue/late fraction), while
at higher redshifts the LFs are consistent with no evolution. A
comparison with our hierarchical CDM model shows a good agreement at
bright and intermediate magnitudes. A better agreement of the model
has also been found at fainter magnitudes due to the suppression of
star formation in small objects by the action of an ionising UV
background.

The shape of the red/early luminosity function is better constrained
only at low and intermediate redshifts and it shows an excess of
faint red dwarfs with respect to the extrapolation of a flat
Schechter function. In fact a minimum around magnitude $M_B(AB)=-18$
is present together with an upturn at fainter magnitudes. This
peculiar shape has been represented by the sum of two Schechter
functions.

We found that the bright one is constant up to $z\sim 0.7$ beyond
which it decreases in density by a factor $\sim 5$ (10 for the early
galaxies) up to redshift $z\sim 3.5$. The comparison with our
hierarchical CDM model shows that, although the predicted LF has a
slight flattening at intermediate luminosity, the model still
overpredicts the LF at faint magnitudes. The bright end of blue and
red LFs at low and intermediate  redshifts is  in   good  agreement
with  recent estimates  from   the  DEEP2 spectroscopic survey. As a
consequence  of   this  complex  evolutionary  behaviour, the
luminosity  densities  of  the  relatively   bright
($M_B(AB)<-20.2$) red/early and  blue/late galaxies  show a
bifurcation beyond redshift $z\sim  1$. Indeed  the  LD of  the
blue/late  population  keeps  increasing  up  to    $z\sim  3.5$,
while the   luminosity density  of red/early galaxies decreases  by
a factor  $\sim 2-3$ respectively  in the $z=1-3.5$ interval.

To derive  hints on  the nature  of the  galaxies responsible  for
the peculiar shape of the red/early  LF, we have performed an
analysis of their stellar masses and spatial distribution. We found
that the early galaxies have systematically higher stellar masses
with respect to the late ones for a given B band luminosity.
Brighter early galaxies have  a spatial distribution  more
concentrated  in higher  density regions if compared  to  the late
ones  of the  same  luminosity class.  On  the contrary, fainter
early and late  galaxies  show a very similar spatial distribution.
Thus, the  different environmental properties  do not seem  to  be
the  main   responsible  for the difference  in shape at
intermediate magnitudes between the blue and red  LFs. The latter
seem  to   stem   from  the different star formation and feedback
histories corresponding to different possible merging trees
(evolutionary paths)  leading  to the final assembled galaxy; this
specific history, driving  the evolution of the   star formation,
leads to the different  $M/L$ ratios characterising the different
properties  of blue/late  and red/early galaxies. In summary, the
peculiar shape of the red LF is mainly driven by the {\it  nature}
of the galaxy merging tree rather than by the {\it nurture} where
the galaxy has grown.

\begin{acknowledgements}
We thank the anonymous referee for his/her helpful comments, that
led to a significant improvement of the paper. We thank the whole
GOODS Team for providing all the imaging material available
worldwide. Observations have been carried out using the Very Large
Telescope at the ESO Paranal Observatory under Program IDs
LP168.A-0485 and ID 170.A-0788 and the ESO Science Archive under
Program IDs 64.O-0643, 66.A-0572, 68.A-0544, 164.O-0561, 163.N-0210
and 60.A-9120.
\end{acknowledgements}

\bibliographystyle{aa}
\bibliography{salimbeni}
\end{document}